# A new DNA alignment method based on inverted index

Wang Liang, Zhao KaiYong

Tencent Tech, 100080, P.R. China,

Hong Kong Baptist University, Department of Computer Science, HK, P.R. China

*To whom correspondence should be addressed. E-mail:wangliang.f@gmail.com

[**Abstract**] This paper presents a novel DNA sequences alignment method based on inverted index. Now most large scale information retrieval system are all use inverted index as the basic data structure. But its application in DNA sequence alignment is still not found. This paper just discuss such applications. Three main problems, DNA segmenting, long DNA query search, DNA search ranking algorithm and evaluation method are detailed respectively. This research presents a new avenue to build more effective DNA alignment methods.
[**Keywords**] DNA search engine, BLAST, inverted index

## 1 Introduction

Looking for the faster DNA alignment and analyzing method is always the aim of biologists, especially when the cost of human Genome sequencing reaches $1000. But alignment speed of most current algorithms like BLAST is relatively very slow comparing with the exploding increment of DNA sequences. For example, the number of full genome sequence has reach thousands level, find a sequence in all this genome is still a very touch mission. But if you want to search a word sequence in billions of documents in Internet, you only need several ms using Google. So could we build a DNA search engine like Google? This paper just solves this problem.

Now most DNA search and comparing methods are similar to BLAST/FASTA algorithm, which compares one sequence with the other sequences on by one [1,2]. Faster hash-table based heuristic methods like BLAT [3] and SSAHA [4] are also proposed. The suffix tree based methods are also used to align large genomic sequences [5]. Although many heuristic and pre-index methods could greatly reduce the search time, it's still difficult to meet the challenge in this DNA information explosion period.

In fact, Search engine and DNA alignment solve one same problem, string matching problem. Most algorithms like suffix tree are also shared in these two areas. According to the excellent performance in web pages 'alignment', we believe search engine methods could also provide promising solution for DNA alignment. But its application in DNA analyzing is still not found. We could compare BLAST and search engine methods as follows:

(1) Their main aim are all find the "similar" sequences in databases with query sequence.

(2) The definition of "similar" are almost same. Blast normally consider the number same letters and gaps. Search engine also use the number of same words, distance between to calculate such "similarity".

(3) Comparing very long sequence or multiple Comparing normally apply approved version algorithm based on basic method.

(4) They also have the similar evaluation metrics. Blast apply sensitivity and selectivity to evaluate alignment algorithm. Correspondingly, search engine use recall and precision. Such evaluation metrics are almost same.

(5) The differences: Blast compares the sequences by letters, but search engine use "words". Moreover, Blast and related improvement versions use position to index letters. Search engine index 'document', normally a short sequence. For long sequence, they divide the long text into short "document". So the result of search engine are a series of documents. Results of Blast are segments of sequences. But search engine result show form is normally the dynamic abstract of document, which is also the combination of match segments of result document.

Because the short sequence alignment is the foundation for other kinds of alignments, we mainly discuss the short query, normally 100+bps DNA sequence (read length of most current sequencing methods) alignment in this paper.

The following section will discuss how to use search engine methods to index DNA data, include DNA document construction, long DNA query retrieval, and DNA search ranking algorithm. Section 3 compare the DNA search engine and BLAST by same evaluation method. The last section is a short summary.

## 2 DNA search engine
### 2.1 DNA documents

Search engine use "document" as the basic index item. When searching, they retrieval related documents according query sequence. Normally, documents is longer than query. For very long sequence like a book, we normally divided into short documents. There is no a strict standard to determine the proper length of "document". To ensure the distances between matched words is not long, and also avoid the missing match in section part of two documents, this length should be ten or more multiples length of most queries.

For Refseq database, their sequence length of single record ranges from100 to 10000. We could directly use its record as 'document'. For full genome, we need divide long sequence into short sequences. Here assume most DNA query is about 100 bps, we divide genomes into 1000 bps pieces of "document".

### 2.2 DNA words

The search engine normally uses he words as the basic align item. It's a basic obstacle to apply the inverted index method.

For some language like English, the sequence is naturally segmented into words by space and punctuation. But for DNA sequence, there is no space or punctuation. In some East-Asia language like Chinese, there are also no natural delimiters. It's also a big problem to deal with these languages by computer.

In these years, this problem have been solved to large extent. The simple method to divide the Chinese is n-grams cross segmentation. For example, we select 3-grams, the sequence will be divided into "Ilo/ lov/ ove/vea …". We could also apply this method to segment the DNA sequence.

Moreover, if we apply this methods, the index of search system will become much large. It's also difficult to deal with the gap problems. So most current East-Asia language search engine apply the dictionary based segment method. Such method will divide the "Iloveapple" into "I/ love/ apple". For DNA sequence, we need build an unsupervised segment method to segment DNA sequence into DNA words [6,7]. In our experiment, we use 12 bps as the maximal length of DNA words.

**2.3 DNA inverted index**

After segmenting the DNA sequence, we could build the invert index, which is the basic data structure for search engine.

We could use a simple example to show the inverted index methods:
Three documents:
Document 0, D0= "ATCG ATT ACC";
Document 1, D1= "ATCG ACG AAA ACC";
Document 2, D2 = "ATT ACG AAA ATTC ACC";

In search engine, the inverted index is the mapping from "word" to "document", which is shown in Table.1, Table.2

Table 1; inverted index structure

| word | Doc id |
|---|---|
| ATCG | D0,D1, |
| ATT | D0,D2 |
| ACC | D0,D1,D2 |
| ACG | D1,D2 |
| AAA | D1,D2 |
| ATTC | D2 |

Table 2; inverted index structure, with word position

| word | (Doc id, word position) |
|---|---|
| ATCG | (D0,0),(D1,0) |
| ATT | (D0,1),(D2,0) |
| ACC | (D0,2),(D1,3),(D2,4) |
| ACG | (D1,1),(D2,1) |
| AAA | (D1,2),(D2,2) |
| ATTC | (D2,3) |

If we want to search "ATCG", we could find the "ATCG" in index and get its corresponding Document D0 and D1, which are search result.

If the query contain more words like "ATCG ACC", we should obtain their corresponding Document ID list and then get their intersection set:
Index('ATCG') -> {D0,D1}
Index('ACC') -> {D0,D1,D2}

The final search result = $\{D0, D1\} \cap \{D0, D1, D2\} = \{D0, D1\}$

So {D0,D1} will be the final search result. Normally, the word position are also stored in inverted index. So we could calculate the gap between match words and get the ranking score of matched document.

**2.4 DNA sequence searching**

The search engine support "OR/AND" operator. Most query length are 1-3 words. The "AND" operation are shown in section above.

The "OR" operation is simpler. They get the union set of Document ID list of all query words. For example:

Query: "ATCG ACC"

Index('ATCG') -> {D0,D1}
Index('ACC') -> {D0,D1,D2}

The final search result = $\{D0, D1\} \cup \{D0, D1, D2\} = \{D0, D1, D2\}$

Because the DNA alignment allow some query words missing in matched sequence, DNA search could be regarded as a kind of "OR" operation. More precisely, DNA short query is about 100+ bps, normally 10+ words, it correspond to the "long query" search in search engine. Such system mainly use "WAND" operator, which means all query words 'should' be in result sequence. It doesn't need all query words appear in result sequence, but the sequence match more words will get higher scores. So it's more appropriate for DNA search.

The WAND operation is very similar to "OR", but it apply the improved document list merging algorithms for "long query" search. "WAND" operation will get more precise search results. Its searching speed is also faster than "OR" operation [8,9]. So we use "WAND" operation as the basic DNA searching algorithm.

**2.5 DNA search result ranking**

Based on inverted index, we could get the candidate result list by inverted list merge operation like WAND. One query may retrieval thousands of documents, so we need sort the candidate results according some rules.

The BLAST normally select several 'high-scoring words' from query sequence. Then it gets the candidate sequences according to this selected words. These sequence is ranked by local alignment smith-waterman score at last.

The search engine also select most 'words' except few 'stops words' from query sequence and then obtain the candidate sequences. Its ranking algorithms is normally TF*IDF, BM25, etc. Gaps of query words could also be regarded as ranking feature. And of course, we could also re-rank its candidate results by BLAST's ranking methods.

## 3  Comparison between blast and search engine method

Here we use an open source search engine Xapian to build the experimental system. The results ranking algorithm is default BM25. What we need is to input the segmented sequence into search system [10].

We could compare the search engine method and blast as follows. We use full genome of human as experimental data and divide it into 1000 bps length documents. Then we apply BLAST (BLAST+ version) and search engine to index these data respectively. The 10000 queries are built by randomly select 100-130 bps sequence from different genomes data. In these queries, a 1000 queries set are selected from human genome. Normally the results of BLAST is a segment of document, but the search engine is document. To compare their results, we also use the segment's corresponding segment as the return results of BLAST. Because the document is only 1000 bps, this rule is reasonable.

Their search speed are shown inTable3:

Table.3 search speed of BLAST and search engine

|  | average time | Max time | Min time | Memory need |
|---|---|---|---|---|
| BLAST | 385ms | >30s | <1s | >2G |
| Search engine | 425ms | >5s | <50ms | <100M |

From Table.3, we could find the speed of search engine is similar to BLAST, although search engine use Disk but BLAST use memory. Using disk index, we could easily run human genomes search in 100M memory. But BLAST will need more than 2G memory. We could even index many genomes in a personal computer by search engine system. This will give many advantages for researchers.

Moreover, there are also many mature open source systems to build large scale search engine, using these systems, we could easily build a search engine indexing all current DNA sequences. Most search system all contains the disk index and memory index module.

Then we compare their search results. Here we run the exactly matching search test and similar matching test respectively.

For the exact matching test, we wish the query sequence also appear in the search results. The query are segment of 'documents', which are extracted from human genome. If the query's corresponding 'document' appear in search results, we think it pass the exact matching test. Here we use 1000 queries selected from human genome. The test results are shown in Table.4:

Table 4. Exact matching test

|  | exact match(in top 100) | exact match(in top 1000) |
|---|---|---|
| BLAST | 92% | 95% |
| Search Engine | 85% | 90% |

We find BLAST could obtain about 95% exactly match sequences. The search engine also returns 90% exactly match sequences. We could also rewrite the ranking module of search system by BLAST's re-rank methods to get the better performance.

Then we run the similar match test. We use the whole sequence of return 'document' of BLAST and search engine as the object sequences, and then compare them with query sequence by smith-waterman algorithm. Their smith-waterman score are shown in Table.5:

Table 5. Smith-waterman score comparison

|  | Average score (top 1) | Average score (top 10) |
|---|---|---|
| BLAST | 67.4 | 44.8 |
| Search Engine | 68.5 | 62 |

From Table5, we could find the results of search is a litter better than BLAST in sensitivity /recall. This mainly because search engine directly use all words to seek candidate sequences. There is no complex high-scoring words operation. It's very similar to FASTA search method. But the documents based inverted index could also ensure its search speed.

## 4 conclusion

The search engine and sequence alignment almost do the same thing. Based on inverted index of search engine, we give a completely new DNA alignment method, which has some distinct advantages such as low cost, simpler, and even more effective.

The deeply and widely research in search engine area could provide more new ideas for DNA alignment, include search algorithms, distributed search design, etc. For example, some search engines use 'latent semantic index' to ensure the recall rate [11]. Like in Google, query "phone" could return the pages that only contain "GALAX" and no "phone". This feature may be also appropriate to index the DNA sequence.

Moreover, in some databases like Refseq, the combination search request like DNA sequence about "transgene" and similar with "AATTCCAGGG….." are very common. We need the complex design to meet this request. But in DNA search system, the DNA sequence and reference text are indexed by the same way, we could easily apply the multiple field search to realize such combination search. We have built a demo system for it [12].